\documentclass[conference]{IEEEtran}
\usepackage{url}
\usepackage{graphicx}

\ifCLASSINFOpdf
\else
\fi
\IEEEoverridecommandlockouts
\begin{document}
%
\title{Two Decades of SCADA Exploitation:\\A Brief History
\thanks{This is a preprint of a publication published at the 1st IEEE Conference on Application, Information and Network Security (AINS).
Please cite as: S. D. Duque Anton, D. Fraunholz, C. Lipps, F. Pohl, M. Zimmermann, H. D. Schotten, ``Two Decades of SCADA Exploitation: A Brief History,'' in: \textit{2017 IEEE Conference on Application, Information and Network Security (AINS)}. IEEE, IEEE Press, 2017, pp. 98-104.}}

\author{\IEEEauthorblockN{Simon Duque Ant\'{o}n, Daniel Fraunholz, Christoph Lipps,\\ Frederic Pohl,  Marc Zimmermann and Hans D. Schotten}
\IEEEauthorblockA{Intelligent Networks Research Group\\
German Research Center for Artificial Intelligence\\
DE-67663 Kaiserslautern \\
Email: \{firstname\}.\{lastname\}@dfki.de}
}


%


\maketitle

\begin{abstract}
Since the early 1960,
industrial process control has been applied by electric systems.
In the mid 1970's, the term SCADA emerged, describing the automated control and data acquisition.
Since most industrial and automation networks were physically isolated, security was not an issue.
This changed, when in the early 2000's industrial networks were opened to the public internet.
The reasons were manifold.
Increased interconnectivity  led to more productivity,
simplicity and ease of use.
It decreased the configuration overhead and downtimes for system adjustments.
However, it also led to an abundance of new attack vectors.
In recent time, there has been a remarkable amount of attacks on industrial companies and infrastructures.
In this paper, known attacks on industrial systems are analysed. 
This is done by investigating the exploits that are available on public sources.
The different types of attacks and their points of entry are reviewed in this paper.
Trends in exploitation as well as targeted attack campaigns against industrial enterprises are introduced.
\end{abstract}


%
\IEEEpeerreviewmaketitle

\section{Introduction}
In the 1970's, the third industrial revolution took place~\cite{Thomson.2015}.
During this phase, computers were introduced into industry in order to automate tasks that,
until then,
had to be done by hand or by application-tailored solutions.
Since then, the computer technology has taken huge steps.
Reconfigurable Programmable Logic Controllers (\textit{PLCs}) took the place of hard-wired relay logic circuits~\cite{Galloway.2013}.
Domain-specific, proprietary fieldbuses, like \textit{CAN}~\cite{RobertBoschGmbH.1991} and \textit{Modbus}~\cite{MODICONInc..1996,  ModbusIDA.2006},
have been replaced by \textit{TCP/IP}-based solutions,
such as \textit{ModbusTCP}~\cite{ModbusIDA.2006, Modbus.2012}, \textit{ProfiNET}~\cite{PROFIBUS.2017} and \textit{OPC UA}~\cite{OPCFoundation.2017},
that make use of the vastly available internet infrastructure and its network hardware.
Opening networks to the outside enables easier management of production capabilities.
Remote maintenance, simpler adjustment of machines and a constant flow of information are but a few of the advantages.
There are, however, some downsides. 
Two of the main reasons why security is inherently absent in virtually every technology and protocol used, are as follows:
Industrial networks were physically separated from the internet, when the technology arose~\cite{Igure.2006}
and each set up of an industrial company is unique and very hard to get around in~\cite{Igure.2006}.
As recent events,
many of which are explained in section~\ref{sec:attack_campaigns},
show, both assertions do not hold true anymore,
if they ever did.
Many recent examples show that industrial networks can and will be breached.
It needs to be highlighted, that, as in consumer electronics, the user plays a crucial role in securing a system.
Many of the newer botnets,
such as Hajime or Mirai,
try to gain access by using default credentials,
with a tremendous success.
This behaviour has been analysed,
among others,
in our previous works~\cite{Fraunholz.2017, Fraunholz.2017_2}.
Many industrial systems use credentials for means of configuration.
For reasons of ease of use, however,
the passwords are often weak and shared among many users.
Attackers that try standard configurations to gain access will succeed if the system credentials have not been altered.
This kind of threat is also common in the exploits examined in section~\ref{sec:in-depth_anal}.
It is very hard for intrusion detection systems to discover abuse that is performed with valid credentials.
Changing default credentials is therefore a vital step in order to enable security in a system.
The remainder of this work is structured as follows.
In section~\ref{sec:related_work}, surveys and analyses of attacks are listed.
After that, a statistical analysis of the Common Vulnerabilities and Exposures (\textit{CVE}) list is performed in section~\ref{sec:stat_anal}.
This is followed by an in-depth analysis of available Supervisory Control And Data Acquisition (\textit{SCADA})-system based exploits in section~\ref{sec:in-depth_anal},
as well as a breakdown of attack campaigns against industry in section~\ref{sec:attack_campaigns}.
The lessons learned are listed in section~\ref{sec:lessons_learned}.
This work will be concluded in section~\ref{sec:conclusion}.

\section{Related Work}
\label{sec:related_work}
Even though there are a lot of survey papers,
as well as taxonomies that present an overview of different kinds of attacks,
there has not yet been a systematic analysis of all publicly available \textit{SCADA} exploits to the best of our knowledge.
A very broad and extensive overview over current \textit{SCADA}-based attack-vectors can be found in the works of Zhu, Joseph and Sastry~\cite{Zhu.2011}.
In addition to that, there are other works that give an overview over existing SCADA-attacks and survey current exploits~\cite{Igure.2006, MottaPires.2006, Caswell.2011, Meixell.July2013}.
Not only attacks on \textit{SCADA}-systems are well documented, but also countermeasures, as well as means for hardening systems, are processed in literature~\cite{Chandia.2007, 
HildickSmith.2005}.
There are also works presenting taxonomies of attacks, also in order to help operators assess risks and threats to their systems and implement the according countermeasures~\cite{InternationalOrganizationforStandardization.2013, Langfinger.2016}, as well as works for the collection of data that allows for insight about the condition of a system~\cite{DuqueAnton.2017, DuqueAnton_2.2017}.
The German Federal Office for Information Security (\textit{BSI}) periodically releases security advices for industry~\cite{BundesamtfurSicherheitinderInformationstechnik.2016}.
Furthermore, there are surveys analysing specific domains, such as automotive and fieldbus-security~\cite{Checkoway.2011} (some of the relevant works are in German~\cite{Wolf.2014, TSystems.2016}) and wireless-security~\cite{Wright.2007}.	
Many of the exploits we examine in this paper have already been investigated in literature.
The amount of works analysing singular attacks is vast, therefore, we only reference such works in the according sections.

\section{Statistical Analysis}
\label{sec:stat_anal}
An exhaustive list of all \textit{CVEs} can be found online~\cite{MITRE.2016}.
Since it contains over 100 000 entries,
manual analysis was infeasible.
We developed a text-processing script in order to gain statistical information about the distribution of exploits.
A major drawback was that the most specific information was written in natural language, without any form.
We searched the document for keywords while using stemming in order to find any variant of the keyword.
Stemming is a technique employed to process natural languages~\cite{Lovins.1968}.
The word stems of keywords are derived, then similar word stems are searched in the target file.
We used the python stemming-library~\cite{Chaput.2010}.
The results of the statistical analysis are summarised in table~\ref{tab:stat_cve_anal}.

\begin{table*}[!h]
\centering
\caption{Statistical Analysis of the \textit{CVE}-Library}
\label{tab:stat_cve_anal}
\begin{tabular}{| l | c || r | r |}
\hline
Description & Keywords & Number & Percentage \\
\hline
\hline
All CVEs & - & 106 540 & 100.00\% \\
\hline
Remote Code Execution & rce, arbitrary, execution & 28 016 & 26.30\% \\
\hline
Denial of Service & denial, crash, instable, consume & 19 638 & 18.43\% \\
\hline
Injection attacks & injection, sql & 17 280 & 16.22\% \\
\hline
Information Disclosure & traverse, disclose, sensitive, bypass & 14 875 & 13.96\% \\
\hline
Buffer Overflows & buffer, overflow & 9 800 & 9.20\% \\
\hline
SCADA-attacks & scada, plc, industry, modbus, profinet, beckhoff, siemens  & 373 & 0.35\% \\
\hline
\hline
Overall categorized entries & - & 65 919 & 61.87\% \\
\hline
Entries w/ multiple keywords & - & 21 620 & 20.29\%\\
\hline
\end{tabular}
\end{table*}

The entry "Overall categorized entries", as well as the "Percentage covered by keywords",  display the number of different attacks that have been classified, after accounting for entries with multiple keywords. 
That means 65 919 entries (or 61.87\%) in the \textit{CVE} list can be attributed to at least one of the categories.
The largest group is Remote Code Execution with 28 000 occurrences, closely followed by Denial of Service (\textit{DoS}) and Injection attacks.
\textit{SCADA} exploits are relatively small, with only 373 entries. 
This shows that, even though  it is not as present as office \textit{IT}-based attacks,
\textit{SCADA}-based exploits are becoming more of an issue for manufacturers.

\section{In-depth Analysis}
\label{sec:in-depth_anal}
In this section,
four different types of attacks that are relevant for industrial applications are analysed.
First, attacks on \textit{PLC} systems are considered in subsection~\ref{ssec:plc}.
After that, fieldbus-based exploits are discussed in subsection~\ref{ssec:fieldbus}, followed by wireless- and hardware-attacks in subsections~\ref{ssec:wireless} and~\ref{ssec:hw}.
These types of attacks were chosen to be discussed as they are the industrial-specific attack vectors and have not be discussed at large in the context of office-\textit{IT}-security.
\textit{PLCs} can mostly be found in industrial environments as they are used to control production machines.
The same goes for fieldbus systems,
that,
aside from some appliances in home automation,
are comonly employed in industrial automation.
Wireless networks are also commonly used in office and home environments.
There are,
however,
industry specific protocols that are only applied in this context.
These protocols are discussed here.
Hardware attacks can have a great impact due to the distributed nature of production environment and the fact that machines have hardware interfaces.

\subsection{Attacks on PLCs}
\label{ssec:plc}
\begin{figure}[!h]
\centering
\includegraphics[width=0.45\textwidth]{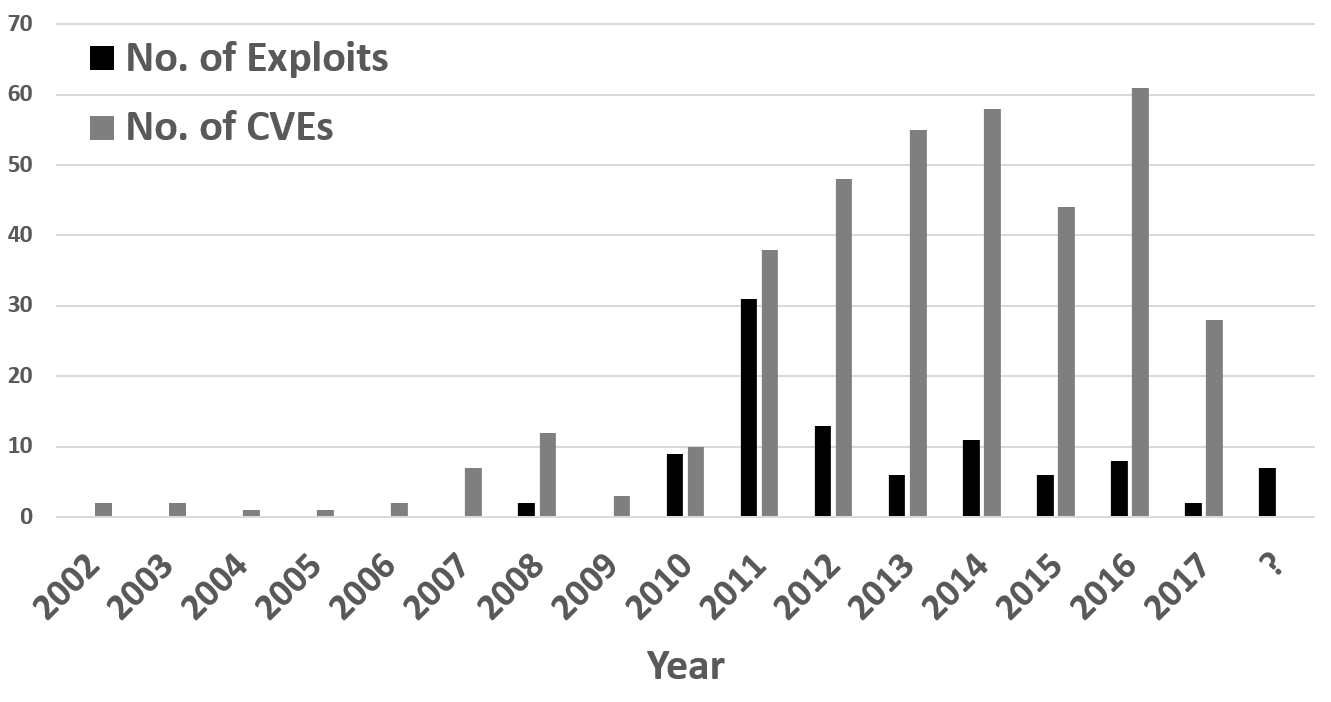}
\caption{Number of Exploit and \textit{CVE} Discoveries per Year}
\label{fig:exp_per_year}
\end{figure}

\textit{PLCs} are resource for industrial applications controlling Cyber-Physical (Production) Systems.
Hence,
they interact with and operate devices in the physical world.
In contrast to office \textit{IT} systems which only handle data,
they interact with the real world.
Attacks on \textit{PLCs} therefore have an impact on physical entites,
be it human workers or production resources.
This leads to grave consequences of the successful abuse of \textit{PLCs}.
As common computation resources,
\textit{PLCs} usually require an underlying operating system.
In most cases,
this is a version of Windows,
adapted to the specific needs for industrial applications.
As there is an abundance of exploits and vulnerabilites based on flaws in the operating system,
we only consider vulnerabilities that specifically derive from the industrial application of the given system.
Furthermore, 
only threats that occur in this context are analysed.
In total, we found about 100 exploits as \textit{metasploit}~\cite{Rapid7.2010} modules and  Proofs of Concepts (\textit{PoC}).
All metasploit-modules are listed in the \textit{Rapid7}-database~\cite{Rapid7.2000}.
The databases we searched additionally were \textit{exploit-db}~\cite{OffensiveSecurity.2009}, \textit{0day-today}~\cite{Inj3ct0rTeam.2008} and \textit{packetstorm-security}~\cite{PacketStormSecurity.1998}.
This number is smaller than the entries found in the \textit{CVE} list in section~\ref{sec:stat_anal} as there is executable code to be found.
As a result, anybody can exploit these vulnerabilities without much difficulties, rendering them very dangerous for operators.
The number of \textit{CVE} discoveries and exploit developments per year is shown in figure~\ref{fig:exp_per_year}.
Unfortunately, some exploits could not be attributed to a year; this has been accounted for by a question mark.
The list amounts to a mean value of 8.8 and a median of 7 exploit developments per year.
A peak of 31 developments per year can be found in 2011. 
One possible explanation is that it was the year after \textit{Stuxnet}~\cite{Falliere2011-stux} was discovered (see table~\ref{tab:campaigns_table}) and there was a special interest in \textit{PLC}-exploitation.
The trend of \textit{CVE}-development is also rising, meaning that the amount of \textit{CVEs} discovered per year has been rising, starting in 2011.

We distinguished between four different categories of exploits:

\begin{itemize}
\item \textit{Code Execution} is the unauthorised execution of malicious code
\item \textit{Data Extraction} is the unauthorised disclosure of information
\item \textit{DoS} describes the partial or full degradation of the availability of a service or resource
\item \textit{Privilege Escalation} is the process of maliciously obtaining higher privileges on a system than intended
\end{itemize}

\begin{figure}[!h]
\centering
\includegraphics[width=0.45\textwidth]{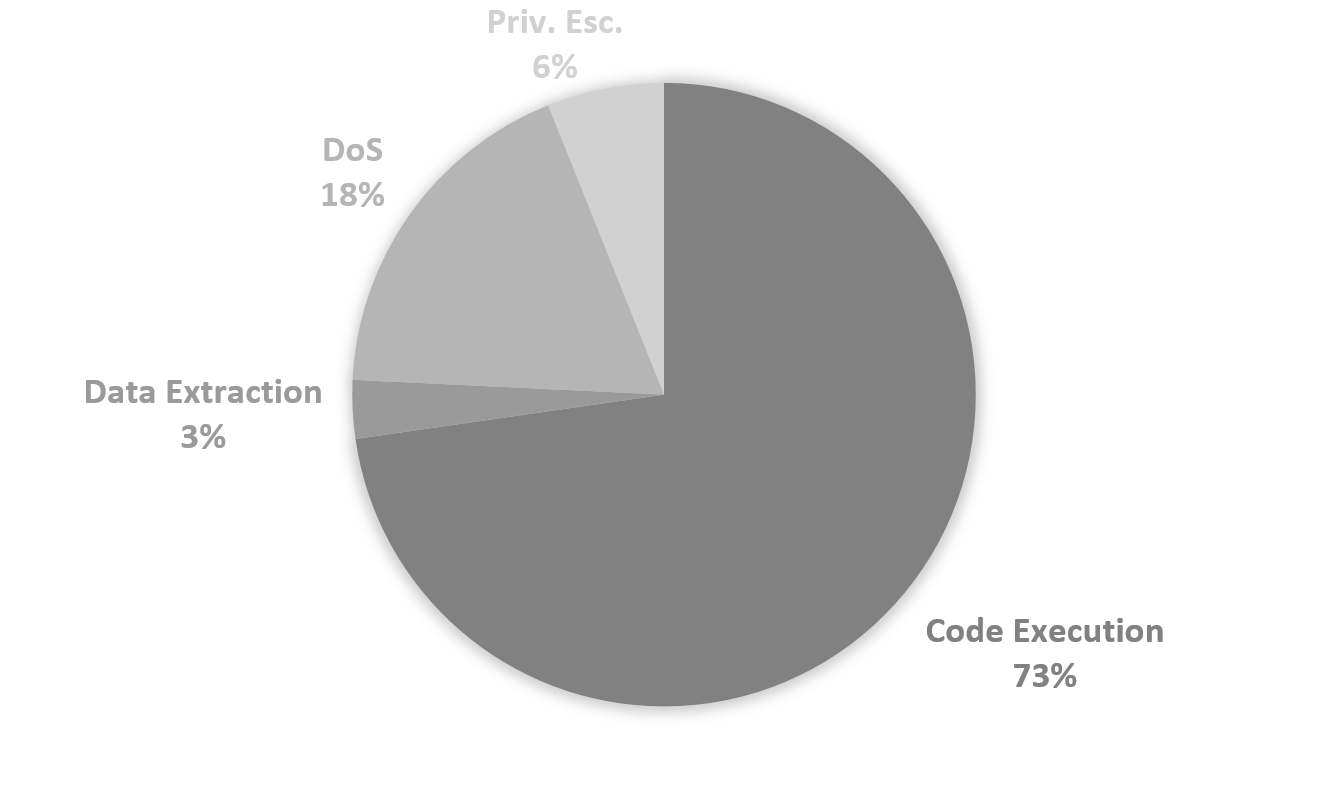}
\caption{Distribution of Categories on Windows Platforms}
\label{fig:windows_cat_per_plat}
\end{figure}

The distribution of these categories on windows-based systems is depicted in figure~\ref{fig:windows_cat_per_plat}.
Of 66 windows-based exploits, almost three quarters allow the execution of arbitrary code.
This is a tremendous threat since it allows an attacker to alter, add and delete resources on the affected system.

\begin{figure}[!h]
\centering
\includegraphics[width=0.45\textwidth]{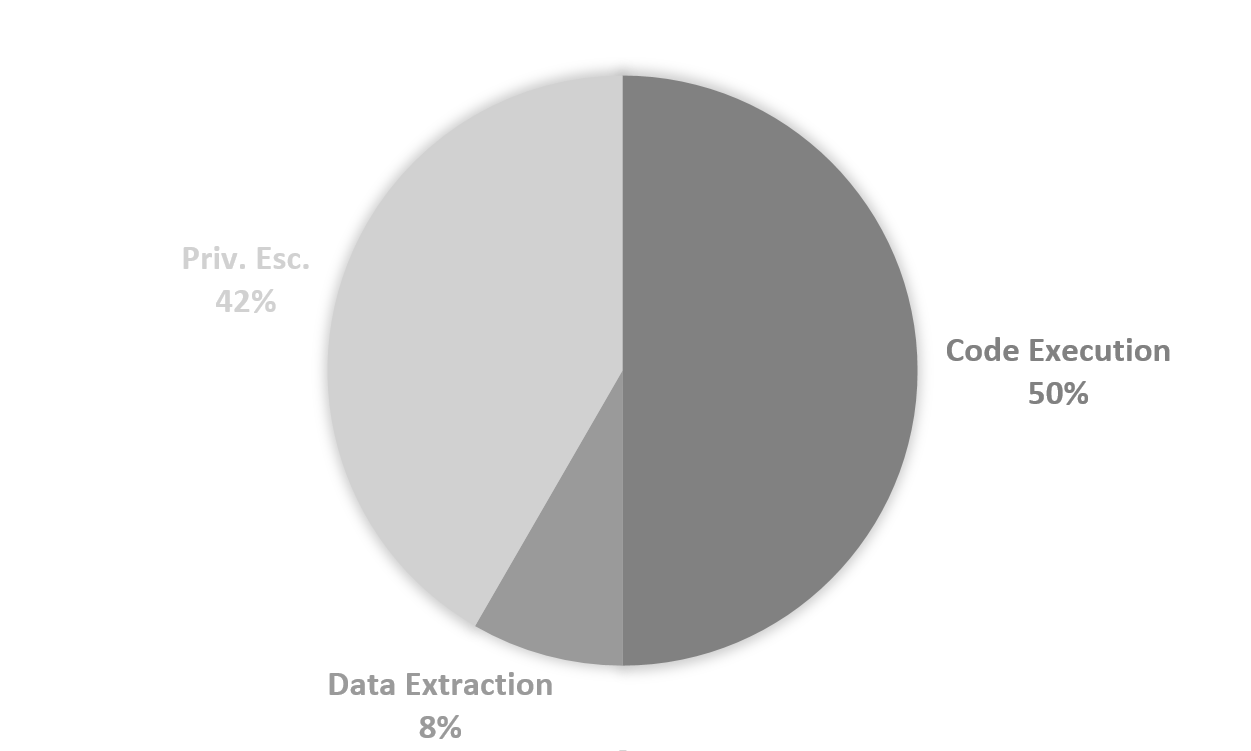}
\caption{Distribution of Categories for Local Exploits}
\label{fig:local_cat_per_acc}
\end{figure}

\begin{figure}[!h]
\centering
\includegraphics[width=0.45\textwidth]{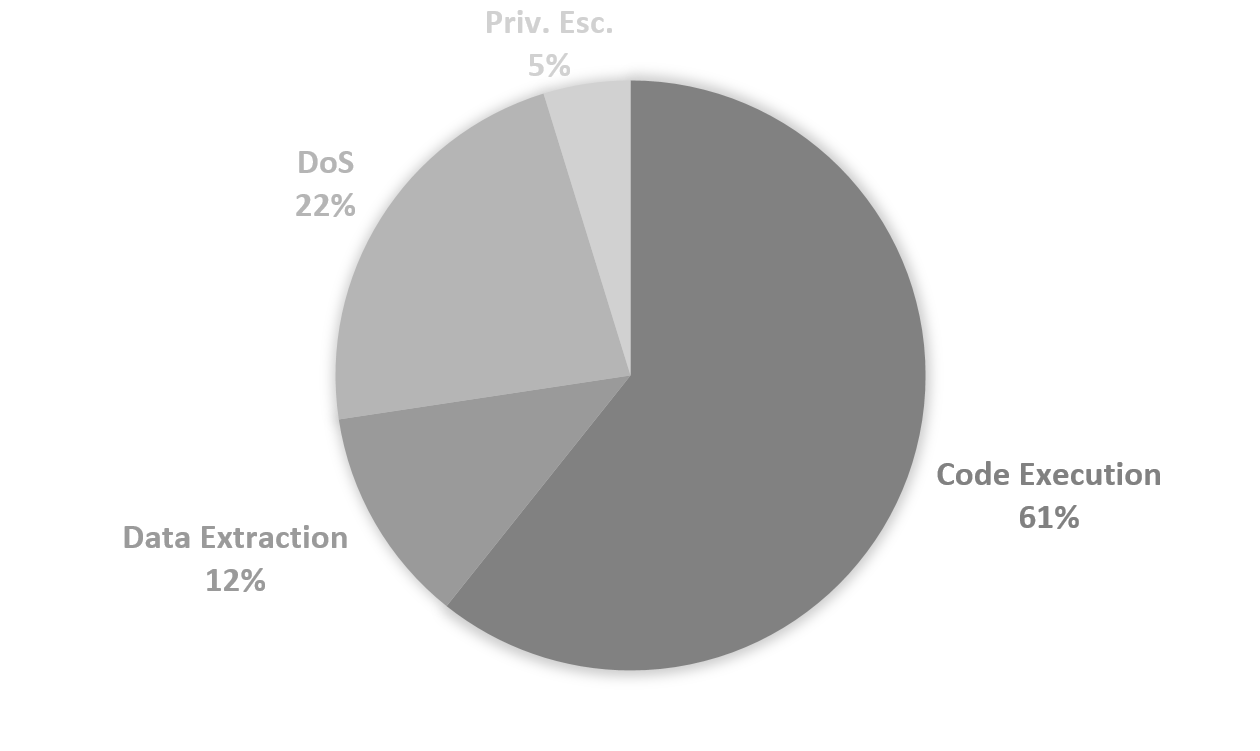}
\caption{Distribution of Categories for Remote Exploits}
\label{fig:remote_cat_per_acc}
\end{figure}

Furthermore, we grouped all exploits into \textit{remote} and \textit{local}.
\textit{Local} exploits allow an attacker to execute an exploit on a system he already has unprivileged access to, usually in the form of a user account with limited rights. 
\textit{Remote} exploits can be executed without any prior access to the system, despite some form of network connection.
In figure~\ref{fig:local_cat_per_acc}, the distribution of the categories for local access is shown. 
The overall number of local exploits is relatively small, comprising only 12 exploits.
In this scenario, the execution of code is most common.
The distribution of the categories for remote access is shown in figure~\ref{fig:remote_cat_per_acc}.
It comprises of 84 exploits, most of which are code execution as well.
The most prevalent threat for \textit{PLC}-based exploitation is the execution of remote code. 
This is a very severe threat because of the priorities of industry.
While in classic office-\textit{IT}, the \textit{CIA} (Confidentiality, Integrity, Availability) security targets are common, each with about the same importance,
the most important security target by far for industry is availability.
Unavailable production facilities cost a huge amount of money, making this the top priority of machine operators.
\textit{Code Execution} has the potential to disable facilities, rendering them unavailable and costing revenue.

\subsection{Attacks on Fieldbus-Level}
\label{ssec:fieldbus}
Due to the proprietary nature of industrial networks, a vast landscape of fieldbus protocols has emerged. 
Protocols such as \textit{Modbus}~\cite{MODICONInc..1996}, \textit{Profinet}~\cite{PROFIBUS.2017}, \textit{CAN}~\cite{RobertBoschGmbH.1991},
\textit{Local Interconnect Network (LIN)}~\cite{LINConsortium.2010}, \textit{Media Oriented System Transport (MOST)}~\cite{MOSTCooperation.2010} and
\textit{FlexRay}~\cite{FlexRayConsortium.2010}.
These protocols have inherent security flaws.
Since there are no means of authentication, identities are not assigned to the participating entities~\cite{Zhu.2011}.
That means an attacker with access to the bus can appear as a valid communication partner and thus extract and inject messages.
This results in a break of confidentiality and integrity.
Due to these security flaws and the lack of encryption~\cite{Porros.2010}, an attacker can monitor the systems and even deploy attacks.
Examples for such attacks are Man in the Middle (\textit{MitM}) and \textit{DoS}.
In systems using \textit{Modbus}, malicious adversaries can read all messages to discover active controllers and used function codes as well as inject commands themselves.
Additionally, they can send incorrect messages or error flags to eliminate single controllers or even the entire system.
Many industrial systems have a remote maintenance interface that can be accessed via internet~\cite{Caswell.2011}.
Often, this interface is secured poorly, or not at all~\cite{Caswell.2011}.
This means that an attacker with access to the same network as the interface can change system settings and read system conditions.
Gateways are used in order to connect several fieldbus networks.
Oftentimes, these gateways are not configured securely, allowing an attacker that has access to one fieldbus network, to traverse to different networks~\cite{Wolf.2014}.
As a counter example, \textit{OPC-UA}~\cite{OPCFoundation.2017} needs to be mentioned.
It is a very modern fieldbus-protocol that allows definition of entities, including authentication and encryption.
The shell model allows for encapsulation of functional units and the definition of interfaces.

\subsection{Attacks on Wireless Systems}
\label{ssec:wireless}Driven by the fourth industrial revolution, wireless communication finds its way into industrial systems.
There are some protocols that are commonly used in industrial applications, 
such as \textit{Bluetooth Low Energy}~\cite{BluetoothSIG.2010}, \textit{ZigBee}~\cite{ZigBeeAlliance.2004} 
and \textit{Z-Wave}~\cite{ABR.2016}, \textit{Radio Frequency IDentifier (RFID)}~\cite{etsi.2017} and the \textit{Long Range Wide Area Network (LoRa)}~\cite{Sornin.2015}.
\textit{Wireless Local Area Network (WLAN)}~\cite{IEEEComputerSociety.2016} is also often used in industry,
but since it was originally developed for classical office-\textit{IT}, it is not considered in this work.
\textit{RFID} is commonly used by industry to tag entities and materials and account for them in storage or production.
The other protocols are commonly used for data transmission and communication.
There are several flaws and fixes for \textit{WLAN}, but they are out of scope for this work for the reasons named above.
As there is no physical access control to the wireless channel, an adversary can listen to the communication, given he is within the range of the wireless signal.
Therefore, most wireless communication protocols are encrypted.
Still, some encryption schemes can be broken, rendering the content unprotected.
If there is no, or weak, encryption, an attacker can listen to the communication and extract information to perform a \textit{MitM}~\cite{Conti.2016} attack.
Furthermore, he can inject messages into the network with the purpose of launching \textit{DoS} attacks.
A famous example is \textit{Wireless Equivalent Privacy (WEP)}~\cite{IEEE802.11.1994}, that is broken~\cite{Fluhrer.2001} but still in use.
Another example is \textit{ZigBee} whose encryption key, in its default configuration, can easily be recovered by an attacker.
Due to poor manufacturer implementations, the secret key is often transmitted in plain text if a new device advertises to the network, for example after restarting~\cite{Zillner.2015}.
An attacker can obtain this key and gains full access to the network.
Another problem in wireless networks are relay attacks.
Using those, an attacker can capture a communication packet, transport it over a different protocol, and inject it into the network on a different place.
This is commonly done with \textit{Bluetooth} or \textit{RFID}.
An attacker can use this method to get a response to a challenge, even though the key is not near a key reader.
This method has already successfully been applied to break the \textit{Passive Keyless Entry and Start (PKES)} of different car manufacturers~\cite{Francillion.2010}.
Spoofing and impersonation are other common attack concepts on wireless protocols. 
Spoofing means the disguise of an attacker as a valid entity to participate in a communication, impersonation describes an attacker that claims to be an entity she is not.
\textit{Bluetooth} is vulnerable to attacks with \textit{Rogue Access Points (APs)}~\cite{Wright.2007}, among others.
Those are \textit{APs} that are set up by an attacker and imitate valid APs.
Because of the ad-hoc nature and the frequency hopping properties of \textit{Bluetooth}, rogue \textit{APs} are hard to detect~\cite{Wright.2007}.
The same concept can be applied to \textit{RFID}, where fake tags or readers can read or manipulate entries~\cite{Garfinkel.2005}.
Furthermore, wireless channels are inherently prone to jamming attacks.
Since there is no access control, an attacker can flood the channel with packets, or simply jam it with noise~\cite{Xu.2005}.
This prevents the valid users from communicating with each other. 
There are also more sophisticated approaches that exploit protocol flaws to prevent communication or that do not jam constantly to make discovery harder~\cite{Xu.2005}.

\subsection{Physical-Layer Attacks}
\label{ssec:hw}
Physical, or hardware attacks, are among the most difficult ones.
An adversary with physical access to a device or system has more possibilities of inflicting damage and abusing services than one on a remote location.
Industrial companies, therefore, put a strong emphasis on obstruction of physical access by perimeters such as, walls, gates and guards.
Given access, an adversary can, with enough force, always destroy a system rendering it unusable and creating a \textit{DoS}.
There are, however, more sophisticated and subtle approaches in tampering with devices.
There are attacks on embedded devices, particularly \textit{PLCs}, that falsify sensor values.
This, in turn, creates, inapt reactions from the devices, leading to undesired behaviour.
In literature, there is the "Ghost in the PLC"-attack, that alters the input-pins of a \textit{PLC}, as described by Abbasi and Hashemi~\cite{Abbasi.2016}.
Another work on falsifying input values and creating improper responses from the system is shown by Urbina, Giraldo, Tippenhauer and Cardenas~\cite{Urbina.2016}.
In addition to tampering with sensor-values, an attacker can read or update the code on a \textit{PLC}.
Such an attack is described by Basnight, Butts, Lopez and Dube~\cite{Basnight.2013}.
In order to stealthily deploy malware on a
\textit{PLC}, Garcia, Brasser, Cintuglu, Sadeghi, Mohammed and Zonouz propose a method to read system information and create a fitting rootkit~\cite{Garcia.2017}.
Even though it is not the most relevant attack vector in practice,
securing physical access is a vital task for industry,
since adversaries with direct access have many opportunities with a potentially high impact.

\section{Attack Campaigns}
\label{sec:attack_campaigns}

The exploits that have been introduced in section~\ref{sec:in-depth_anal} have been used for attack campaigns against industrial players.
We found that there were two noteworthy kinds of attacks:
\begin{itemize}
\item Spearphishing campaigns against employees
\item Attacks on the industrial infrastructure
\end{itemize}

\begin{figure}[!h]
\centering
\includegraphics[width=0.45\textwidth]{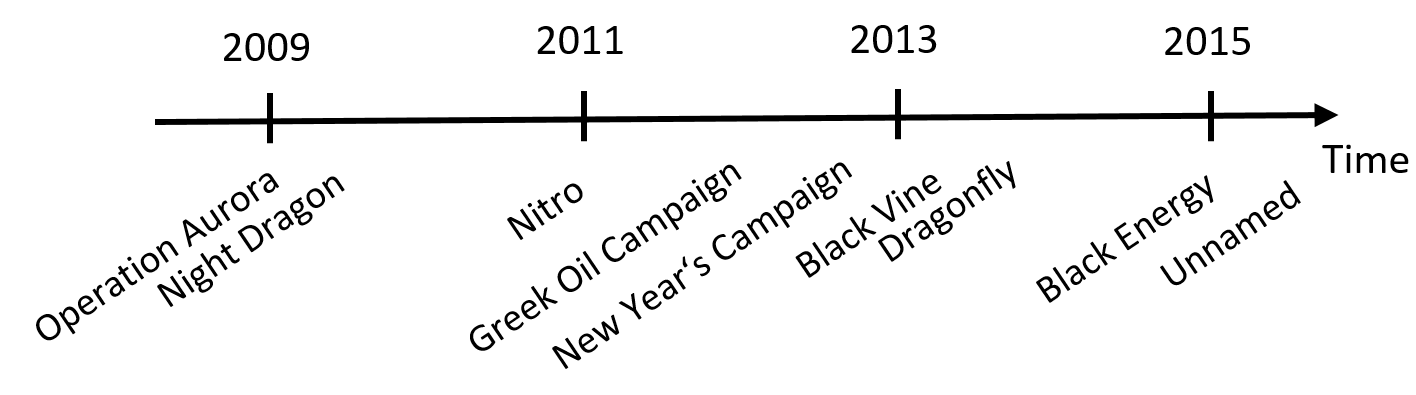}
\caption{Timeline of Selected Spearphishing Campaigns}
\label{fig:timeline}
\end{figure}

Phishing and spearphishing are common practices for malicious adversaries intending to gain insight on company secrets by gaining access to the office \textit{IT} infrastructure and stealing data.
A timeline of known spearphishing campaigns with an industrial background is shown in figure~\ref{fig:timeline}.
In phishing, unsuspecting victims are sent emails with malicious content, oftentimes a link to a website that is infected with malware~\cite{Wood2016-threats}.
Attachments with malicious content are another common form of phishing~\cite{Wood2016-threats}.
The chances of an attacker to get a victim to follow the link can be increased by personalizing the email.
This is called ``social engineering''~\cite{Wood2016-threats},
the application of phishing to selected targets with highly adapted content is called ``spearphishing''.

\textit{Operation Aurora}~\cite{McClure2010-aurora} aimed at the software industry, particularly \textit{Google}.
The \textit{Night Dragon}, \textit{Greek Oil} and \textit{New Year's} campaigns aimed at various branches of the energy industry, namely research and petroleum
processing~\cite{Wueest2014-energy}.
Furthermore, the \textit{Nitro} campaign~\cite{Chien2011-nitro} aimed at the chemical industry and was intended to obtain sensitive documents, designs and schemas for manufacturing.
 \textit{Black Vine}~\cite{DiMaggio2015-blackvine} campaign was used for several targets. First, aerospace companies were in the focus. After that, it was aimed against healthcare institutions in the U.S.
The \textit{Dragonfly}~\cite{secresp2014-dragon} and \textit{Black Energy}~\cite{Lee2016analysis} campaigns aimed at the energy industry as well, this time against
\textit{Industrial Control System (ICS)} manufacturing and power generation.
In a report, an attack campaign, that is called \textit{Unnamed}~\cite{klab2017-threats} in our timeline in figure~\ref{fig:timeline},
was described also aimed for the extraction of confidential information about \textit{ICS} manufacturing in the energy industry.
Attacks on the industrial infrastructure often aim at sabotaging production.
Highly sophisticated malware is employed in these campaigns~\cite{Wood2016-threats}.
A selected list of all known industrial malware campaigns can be found in table~\ref{tab:campaigns_table}.
In this table, the name of the malware is shown, as well  as the year of discovery. 
Furthermore, the presumed target is listed, followed by a \textit{Target Score (TS)} describing the kind of attack that was employed.
The \textit{TS} is assigned a value according to the following scheme:
\begin{itemize}
\item $1$: The malware does not specifically target \textit{ICS}, the incurred consequences are a side effect
\item $2$: The malware targets \textit{Windows} machines related to \textit{ICS}
\item $3$: The malware targets software related to \textit{ICS} projects
\item $4$: The malware targets \textit{PLCs} and other native devices and protocols
\end{itemize}

In addition to that, the presumed purpose, the affected \textit{ICS} and \textit{CVEs} that were used in the exploit are listed.
\textit{Slammer} and \textit{Conficker} were computer worms that also infected nuclear power station~\cite{Kesler2011Vulnerability}
respectively air force stations in France and Germany~\cite{Sciacco2015-air}. 
\textit{Stuxnet}~\cite{Falliere2011-stux} is one of the most renowned industrial malwares. It was aimed at Iranian nuclear enrichment facilities, but, due to programming errors,
also infected other systems and therefore was found.
It used several different 0-day exploits, depending on the operating systems it encountered, and showed a deep understanding of \textit{Siemens S7-300 PLCs}.
\textit{Duqu} and \textit{Duqu 2.0}~\cite{Bencsath2011duqu,Bencsath2015duqu2} were used for spying on industrial project documents. 
\textit{Shamoon} and \textit{Shamoon 2.0}~\cite{Raiu2017Shamoon} were intended on sabotaging the Saudi-Arabian oil industry.
\textit{Stuxnet 0.5}~\cite{McDonald2013stuxnet05} was aimed at sabotaging Iranian nuclear enrichment facilities, also by infecting \textit{Siemens S7-300 PLCs}.
It was employed before \textit{Stuxnet}, but was found later due to a different propagation mechanism.
\textit{Havex}~\cite{secresp2014-dragon} was a malware infecting the European energy industry and spying on confidential information.
\textit{BlackEnergy} and \textit{Industroyer}~\cite{Cherepanov2017} were aimed at Ukrainian power plants.
Major blackouts in December of 2015, respectively December of 2016 in the Ukraine are said to result from \textit{BlackEnergy} and \textit{Industroyer}.

\begin{table*}
\centering
\caption{A Selection of Attack Tools and Campaigns}
\label{tab:campaigns_table}
\begin{tabular}{| l | r | l | r | l | l | l |}
\hline
Name & Year & Presumed Target & \textit{TS} &  Purp. & Affected \textit{ICS} & Exploited \textit{CVE} \\
\hline
\hline
Slammer & 2003 & untargeted & 1 & Sabot. & Nuclear Power Station  & CVE-2002-0649 \\
\hline
Conficker & 2009 & untargeted & 1 & Sabot. & French \& German Air Force  & CVE-2008-4250 \\
\hline
Stuxnet & 2010 & Iranian Nuclear Enrichment Facilites & 4 & Sabot. & Siemens S7-300 & CVE-2010-2568 \\
& & & & & & CVE-2008-4250\\
& & & & & & CVE-2010-2729\\
& & & & & & CVE-2010-2772\\
\hline
Duqu /  Duqu 2.0 & 2011/2015  & Industrial Project Documents & 3 &  Esp. & - & - \\
\hline
Shamoon / Shamoon 2.0 & 2012/2017  & Saudi-Arabian Oil Industry & 2 & Sabot. & - & - \\
\hline
Regin & 2012 & GSM Base Stations & 4 & Esp. & - & -\\
\hline
Stuxnet 0.5 & 2013 & Iranian Nuclear Enrichment Facilites & 4 & Sabot. & Siemens  S7-300 & CVE-2012-3015 \\
\hline
Havex & 2013 & European Energy Industry & 3 &  Esp. & - & -\\
\hline
BlackEnergy & 2016 & Ukrainian Power Plant & 3 & Sabot. & - & CVE-2014-4114 \\
& & & & & & CVE-2014-0751\\
\hline
Industroyer & 2017 & Ukrainian Power Plant & 4 & Sabot. & Siemens SIPROTEC & CVE-2015-5374 \\
\hline
\end{tabular}
\end{table*}

\section{Lessons Learned}
\label{sec:lessons_learned}
We used \textit{Shodan}~\cite{Shodan.}, an internet search engine that specialises on the \textit{Internet of Things (IoT)} and industrial applications.
Specifically, we grouped our search by ports and only looked for ports that are the default for several industrial protocols.
The results of this survey is shown in table~\ref{tab:shodan_table}.
It can be seen that there still is a huge amount of industrial devices to be found, directly connected to the internet.
Since all of the entries in table~\ref{tab:shodan_table} are fieldbuses,
their connection to the internet is risky.
They were never designed for security as one of the paradigms in their development was the physical separation of industrial network and internet~\cite{Igure.2006}.
This assumption does not hold for about 1.45 million fieldbuses, that, depending on their configuration, can be accessed - and probably tampered with - by an attacker via internet access.
We introduced some concepts for botnets in our previous works~\cite{Fraunholz.2017, Fraunholz.2017_2},
and there are other projects that develop industrial honeypots, such as the \textit{Conpot}~\cite{Rist.}-project and the \textit{IoT-pot}~\cite{PaPa.2015}.
One could assume that some of the entries in table~\ref{tab:shodan_table} originate in honeypots.
We found that $137$ of the above entries definitely stem from honeypots by comparing the banners found with the default banners of \textit{Conpot}.
Even though it is plausible that we missed several honeypots, 
we deem it probable that a majority of the entries is from productive systems.
Despite the fact that security flaws in industrial applications have been a critical issue for quite some time,
there still are devices and protocols used in insecure ways.

\begin{table}[!h]
\centering
\caption{Devices Found Publicly Addressable by \textit{Shodan}}
\label{tab:shodan_table}
\begin{tabular}{| l | r | r | r |}
\hline
Service & Port Numbers & Hits & Hit Percentage \\
\hline
\hline
EtherNet/IP &  2222 & 1 015 093 & 69.78\% \\
\hline
DNP3 &  20000 & 232 108 & 15.95\% \\
\hline
OMRON &  9600 & 51 911 & 3.57\% \\
\hline
Niagara Fox &  1911 & 46 806 & 3.22\% \\
\hline
ENIP &  44818 & 32 100 & 2.21\% \\
\hline
Proconos &  20547 & 19 761 & 1.36\% \\
\hline
Modbus &  502 & 18 732 & 1.29\% \\
\hline
CoDeSys &  1200, 2455 & 17 667 & 1.21\% \\
\hline
PCWorx &  1962 & 14 949 & 1.03\% \\
\hline
Siemens &  102 & 3368 & 0.23\% \\
\hline
Fieldbus &  1089-1091 & 924 & 0.06\% \\
\hline
Profinet  &  34962-34964 & 809 & 0.06\% \\
\hline
DNP &  19999 & 300 & 0.02\% \\
\hline
EtherCAT &  34980 & 270 & 0.02\% \\
\hline
\hline
Sum &  - & 1 454 798 & 100.00\% \\
\hline
\end{tabular}
\end{table}

\section{Conclusion}
\label{sec:conclusion}
The trend in figure~\ref{fig:exp_per_year} shows that \textit{PLC}-exploitation is becoming more relevant.
At the same time,
our findings in section~\ref{sec:lessons_learned} point out that many operators do not employ their industrial networks in a physically separated way to at least provide basic security.
In this work, we showed that the kill chain for \textit{ICS} is rather easy to use.
There are tools to identify vulnerable systems, as well as databases that contain information about vulnerabilities and sometimes also the corresponding exploits.
This makes it simple also for non tech-savvy people to attack systems and cause damage.
The rising importance of interconnectivity  in industrial applications will lead to an increase in interest of attackers.
As more and more industrial systems become accessible, get more complex software and are remotely configurable,
the number of possibilities for exploitation and intrusion also increases.
Many industrial operators maintain their production units for decades with little or no possibilities for software updates.
This leads to a tremendous danger, as more exploits occur every year.

\section*{Acknowledgments}
This work has been supported by the Federal Ministry of Education and Research of the Federal Republic of Germany (Foerderkennzeichen KIS4ITS0001, IUNO).
The authors alone are responsible for the content of the paper.


%

\bibliography{bibfile}
\bibliographystyle{IEEEtran}



\end{document}